\documentclass[letterpaper,preprint,12pt,1p]{elsarticle}

\journal{Fluid Phase Equilibria}

\begin{document}
\begin{frontmatter}

\title{Use of the attractive hard-core Yukawa interaction for the derivation of the phase diagram of liquid water.}
\author[cie]{M. Robles \corref{cor1}}
\author[unex]{ M.L\'opez de Haro \fnref{fn1}}
\address[cie]{Centro
de Investigaci\'on en Energ\'{\i}a\\
Universidad Nacional Aut\'onoma de M\'exico,\\
A.P. 34, 62580 Temixco, Mor. M\'exico.}
\address[unex]{Departamento de F\'{\i}sica, Universidad de
Extremadura, E-06071 Badajoz, Spain.}
\cortext[cor1]{Corresponding author. E-mail address: mrp@cie.unam.mx (M. Robles)}
\fntext[fn1]{On sabbatical leave from Centro de Investigaci\'on en
Energ\'{\i}a, Universidad Nacional Aut\'onoma de M\'exico
(U.N.A.M.), Temixco, Morelos 62580, M{e}xico.}

\begin{abstract}
The phase diagram of the attractive hard-core Yukawa fluid derived
previously [M. Robles and M. L\'opez de Haro, J. Phys. Chem. C
\textbf{111}, 15957 (2007)] is used to obtain the liquid-vapor
coexistence curve of real water. To this end, the value of the
inverse range parameter of the  intermolecular potential in the
Yukawa fluid is fixed so that the ratio of the density at the
critical point to the liquid density at the triple point in this
model coincides with the same ratio in water. Subsequently, a
(relatively simple) nonlinear rescaling of the temperature is
performed which allows one to obtain the full liquid vapor
coexistence curve of real water in the temperature-density plane
with good accuracy, except close to the triple point. Such rescaling
may be physically interpreted in terms of an effective
temperature-dependent attractive hard-core Yukawa interaction
potential which in turn introduces an extra temperature dependence
in the equation of state. With the addition of a multiplicative
factor to obtain from the model the critical pressure of real water,
the corrected equation of state yields reasonably accurate isotherms
in the liquid phase region except in the vicinity of the critical
isotherm and in the vicinity of the triple point isotherm. The
liquid-vapor coexistence curves in the pressure-temperature and
pressure-density planes are also computed and a possible way to
improve the quantitative agreement with the real data is pointed
out.
\end{abstract}

\begin{keyword}
 Water, Phase diagram, Hard-Core Yukawa, Equation of state.
\end{keyword}
 
\end{frontmatter}

\section{Introduction}
\label{sec0}
Due to its ubiquitous presence and importance both in our everyday
life and in nature, water has been the subject of numerous
experimental and theoretical studies\cite{Reviews}. Yet, despite all
of this work, the full understanding of its properties is still far
from complete and new efforts are called for.

Out of the many interesting and puzzling aspects of the properties
of water, a particular challenging and still open problem is the
determination of the full phase diagram of the system. In fact, a
stringent test of many theoretical models is often their predicted
phase diagram. In this paper, we will consider an aspect of this
problem, namely the possibility of using a relatively simple model
to obtain a sensible description of the liquid phase and the
saturation curve of liquid water. Concerning the fluid region, the
International Association for the Properties of Water and Steam from
time to time issues documents on their recommended formulation for
the thermodynamic properties of ordinary water substance for general
and scientific use, the most recent one\cite{Release} dated in
September 2009 and related to the so called IAPWS Formulation
1995\cite{Wagner}. This formulation consists of a fundamental
equation for the Helmholtz free energy of water which is obtained
from a sophisticated fitting procedure of a wealth of experimental
results. It is claimed to be valid in the fluid region up to
temperatures of 1273 K and pressures up to 1000 MPa and to represent
very well the most accurate experimental data in the single phase
region and on the vapor-liquid phase boundary. On the computational
side, a rather successful model is the TIP4P model\cite{TIP4P} or
its modified version the TIP4P/2005 model\cite{Carlos0}. This is a
simple rigid nonpolarizable model with a Lennard-Jones site on the
oxygen and partial positive charges on the hydrogen atoms. Further,
the negative charge is placed along the bisector of the H-O-H bond
and the parameters in the potential are adjusted to match the
experimental density and the enthalpy of vaporization of liquid
water. The TIP4P/2005 model in particular provides an excellent
description of the coexisting densities along the orthobaric curve
but it fails to yield the correct critical pressure. As a final
piece of information required for the purposes of this work, mention
should be made here of the analytic equation of state of liquid
water introduced by Jeffery and Austin\cite{Jeffery}. Such an
equation of state is based on the principle of corresponding states
of Ihm, Song and Mason\cite{ISM90} and on the free energy of strong
tetrahedral hydrogen bonds proposed by Poole \textit{et
al}\cite{PSGSA94}. It is accurate and probably useful for
engineering calculations but contains many adjustable parameters
which is clearly an undesirable feature.

Recently\cite{JPC2007}, using thermodynamic perturbation theory and
a hard-sphere (HS) fluid as the reference system, we have derived an
approximate equation of state for the attractive hard-core Yukawa
(HCY) fluid and computed the corresponding phase diagram in the
three different thermodynamic planes. This model fluid, which has
received a lot of attention in different contexts,\cite{HCY} is
interesting because in certain well defined limits its interaction
potential has proved to yield thermodynamically equivalent results
to those of the Lennard-Jones potential, the Coulombic interaction,
the HS potential, or the adhesive HS potential. The main asset of
our derivation lies in the fact that the results are analytical and
have a rather simple form. What we want to do here is to take
advantage of the above mentioned properties and use these results to
obtain a reasonable description of liquid water (including the
vapor-liquid coexistence curve) with a simple model. In this way we
want to provide, albeit with limitations, a positive answer to the
question raised in the provocative title (``Can simple models
describe the phase diagram of water?") of a recent paper by Vega
\emph{et al}\cite{Carlos0bis}.

The paper is organized as follows. In order to make it
self-contained, in the next section we provide the equation of state
of the attractive HCY fluid. This is followed in Sect.\ \ref{sec2}
with the description of the procedure to map the results of the
attractive HCY fluid to those of real water in the fluid phase.
Section \ref{sec3} is devoted to the mapping of the liquid-vapor
coexistence curve in the pressure-density and pressure-temperature
planes. The paper is closed in Sect.\ \ref{sec4} with further
discussion and some concluding remarks.

\section{Approximate equation of state for the attractive HCY fluid.}
\label{sec1}

The attractive HCY fluid is a system whose molecules interact via
the pair potential
\begin{equation}
\phi_{{HCY}}\left( r\right) =\left\{
\begin{array}{cc}
\infty , & r\leq \sigma\\
-\epsilon \frac{\sigma}{r} \exp[-z(r-\sigma)/\sigma], & r>\sigma
\end{array}
\right. ,\label{HCYPI}
\end{equation}
where $\sigma$ is the hard-core diameter, $\epsilon$ is the depth of
the potential well at $r=\sigma$, and $z$ is the (dimensionless)
inverse range parameter.

The potential $\phi_{{HCY}}\left( r\right)$ may be split in a
form convenient for the use of the liquid state perturbation theory
as $\phi_{{HCY}}\left( r\right)=\phi_0\left( r\right)+
\phi_1\left( r\right)$, where the reference potential $\phi_0$ will
be a HS pair potential (where the ``effective" diameter $d$ of the
HS reference system will simply coincide with the hard-core
diameter, $\sigma$), and the perturbed part $\phi_1$ is given by
\begin{equation}
\phi_1\left( r\right) =\left\{
\begin{array}{cc}
0 , & r\leq \sigma \\
-\epsilon\frac{\sigma}{r} \exp[-z(r-\sigma)/\sigma], & r>\sigma
\end{array}
\right.  \label{u1}
\end{equation}

Next we consider the usual perturbative expansion
\cite{Zwanzig:1954} for the (excess) Helmholtz free energy
$A^{{ex}}_{{HCY}}$ of this fluid. To first order in $\beta
\equiv 1/k_BT$, with $T$ the absolute temperature and $k_B$ the
Boltzmann constant, its approximate expression reads (for details
see Ref.\ \cite{JPC2007})
\begin{equation}
\frac{A^{{ex}}_{{HCY}}}{N k_B
T}=\frac{A^{{ex}}_{{HS}}}{N k_B T}-12 \eta \beta \epsilon
e^z G(z), \label{AHCY}
\end{equation}
where
\begin{equation}
\frac{A^{{ex}}_{{HS}}}{N k_B T}= \int_{0}^{\eta
}\frac{Z_{{HS}}(\eta')-1}{\eta' } d \eta'
\label{AHCYbis}
\end{equation}
is the (excess) free energy of the HS fluid,
$Z_{{HS}}(\eta)\equiv 1 + 4 \eta g_{{HS}}(\sigma^+)$ is
the compressibility factor of the HS fluid, $\eta=\frac{\pi}{6}\rho
\sigma^3$ is the packing fraction ($\rho$ being the density and $N$
the number of particles), $g_{{HS}}(\sigma^+)$ is the contact
value of the HS radial distribution function and $G(z)\equiv G(t=z)$
with $G(t)={{\mathcal L}}[r g_{{HS}}(r)]$, ${\mathcal L}$
denoting the Laplace transform operator. Here $g_{{HS}}(r)$
stands for the radial distribution function of the HS fluid.

Within the rational function approximation method\cite{Lopez:2007},
$G(t)$ turns out to be be given by
\begin{equation} {G(t)=\frac{t}{12\eta}\frac{1}{1-e^t
\Phi(t)}} \label{Ltrg}
\end{equation}
where $\Phi(t)$ is a rational function of the form
\begin{equation}
{\Phi(t)=\frac{1+S_1 t+S_2 t^2+S_3 t^3 +S_4 t^4}{1+L_1 t+L_2 t^2}}
\label{phi}
\end{equation}
and the coefficients $S_i$ and $L_i$ are the following functions of
the packing fraction $\eta$
\begin{eqnarray}
{\ L_{1}} &=&{\frac{1}{2}\frac{\eta +12\eta L_{2}+2-24\eta
S_{4}}{2\eta +1},}
 \label{4a} \\
{L_{2}}&=&{-3\left( Z_{{HS}}-1\right) S_{4},}   \label{5}\\
{\ S_{1}} &=&{\frac{3}{2}\eta \frac{-1+4L_{2}-8S_{4}}{2\eta +1},}   \label{4b} \\
{S_{2}} &=&-{\frac{1}{2}\frac{-\eta +8\eta
L_{2}+1-2L_{2}-24\eta S_{4}}{2\eta +1},}  \label{4c} \\
{S_{3}} &=&{\frac{1}{12}\frac{2\eta -\eta ^{2}+12\eta
L_{2}(\eta-1)-1-72\eta ^{2}S_{4}}
{\left( 2\eta +1\right) \eta} ,}   \label{4d}\\
{S_{4}}&=&{\frac{1-\eta }{36\eta \left(Z_{{HS}}-1/3\right) }
\left[ 1-\left[ 1+
\frac{Z_{{HS}}-1/3}{Z_{{HS}}-Z_{{PY}}} \left(
\frac{\chi_{{HS}} } {\chi _{{PY}}}-1\right) \right]
^{1/2}\right] ,}
 \label{6}
\end{eqnarray}
with, $Z_{{PY}}=\frac{1+2\eta +3\eta ^{2}}{\left( 1-\eta
\right) ^{2}}$, $\chi _{{PY}}=\frac{\left( 1-\eta \right)
^{4}}{\left( 1+2\eta \right) ^{2}}$ and $\chi_{{HS}}\equiv
\left(\frac{d}{d\eta}[\eta Z_{{HS}}]\right)^{-1}$. Note then
that the only input required to completely determine
$A^{{ex}}_{{HCY}}$ is an expression for $Z_{{HS}}$.
Here we will consider the compressibility factor that follows from
the popular Carnahan-Starling equation of state given by
\cite{Carnahan:1969}
\begin{equation}
Z_{{HS}}(\eta)=\frac{1+\eta +\eta ^{2}-\eta ^{3}}{\left( 1-\eta
\right) ^{3}}, \label{CS}
\end{equation}
which is known to be rather accurate in the stable fluid region.
This in turn implies that in this case
\begin{equation}
\chi_{{HS}}(\eta)=\frac{\left(1-\eta\right)^{4}}{ 1+4 \eta+4
\eta^{2}-4 \eta^{3}+\eta^{4}}. \label{chiCS}
\end{equation}

From Eqs.\ (\ref{AHCY}) -- (\ref{CS}) one may easily derive the
explicit form of the equation of state of the attractive HCY fluid
in terms of the dimensionless variables $\bar{P}=P \sigma^3
/\epsilon$ ($P$ being the pressure), $\bar{\rho}=\rho \sigma^3$ and
$\bar{T}=k_B T/\epsilon$, namely

\begin{equation}
\bar{P}= \bar{\rho} \bar{T}
Z_{{HS}}\left(\frac{\pi}{6}\bar{\rho}\right)- 2 \pi
{\bar{\rho}}^2 e^z \frac{\partial}{\partial \bar{\rho}} \left[
\bar{\rho} G(z) \right],
\label{PHCY}
\end{equation}
where it is understood that the packing fraction $\eta$ appearing in
all the expressions related to $G(z)$ must be replaced by
$\frac{\pi}{6}\bar{\rho}$.

The (approximate) phase diagram of the attractive HCY fluid in the
three thermodynamic planes ($\bar{T}$-$\bar{\rho}$,
$\bar{P}$-$\bar{T}$ and $\bar{P}$-$\bar{\rho}$) was obtained using
Eq.\ (\ref{PHCY}) and compared to simulation data in Ref.\
\cite{JPC2007}. Such a comparison indicated that the
approximate equation of state was a reasonable compromise between
accuracy and simplicity. With this in mind, in the next section we
will try to use the same equation of state to obtain the phase
diagram of water in the stable fluid region.

\section{Scaling for the thermodynamics of water}
\label{sec2}
The length and energy units for the attractive HCY fluid are defined
arbitrarily by the parameters $\sigma$ and $\epsilon$ in the
interaction potential. The parameter $z$  plays the role of an
adjustable free parameter needed in any application of the model. To
map the attractive HCY fluid to the real thermodynamic properties of
water, we have taken the following steps:

\begin{itemize}
 \item To fix the value of $z$ in the way specified below.
 \item To find a non-linear rescaling of the temperature to fit the saturation
branch of the coexistence curve.
 \item To compare the rescaled attractive HCY fluid isotherms with those of real water.
 \item To adjust the value of the critical pressure of the attractive HCY fluid to match the one of real water.
\end{itemize}

\subsection{The choice of the inverse range parameter}

\begin{figure}[t]
\begin{center}
\includegraphics[height=7truecm]{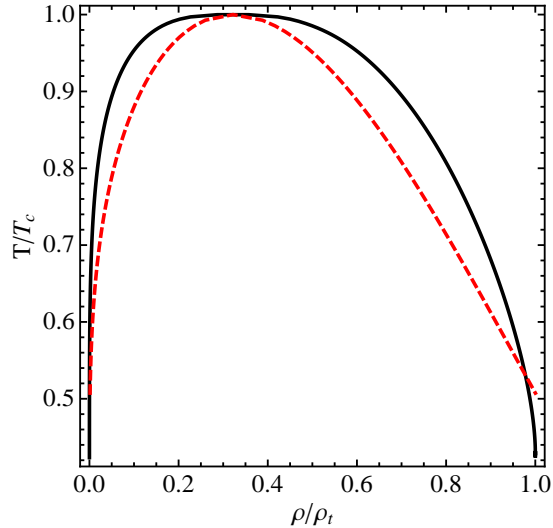} \caption{Coexistence liquid-vapor curves for real
water (continuous line) and the fitted attractive HCY fluid with
$z=1.7118$ (red dashed line).} \label{fig1}
\end{center}
\end{figure}

In our previous work,\cite{JPC2007} we discussed how to obtain, for
a given $z$, the liquid-vapor coexistence curve for the attractive
HCY fluid through the usual Maxwell construction in the
$\bar{T}$-$\bar{\rho}$ plane. Now we want to determine the value of
$z$ such that the coexistence curve derived for the attractive HCY
fluid may be used for water. For this purpose it is convenient to
introduce the reduced quantities $T^*=T/T_c$  and
$\rho^*=\rho/\rho_t$, where $T_c$ and $\rho_t$ are the critical
temperature and liquid density at the triple point, respectively. In
what follows we will add to the subscripts the labels $w$ and $Y$ to
indicate real water and attractive HCY fluid, respectively. For
water, we take $T_{wc}=647.096$ K and $\rho_{wt}=999.793$
kg/m$^3$.\cite{NIST} It should be pointed out that for our
approximate equation of state for the attractive HCY fluid, Eq.\
(\ref{PHCY}), the density at the triple point is independent of $z$
and is given by $\bar{\rho}_{Yt}=0.9375$. \cite{Carlos1} On the
other hand, the critical density and temperature for the attractive
HCY fluid only depend  on $z$. We next fix $z$ by requiring that the
ratio $\rho_{Yc}/\rho_{Yt}$ of the attractive HCY fluid coincides
with the reduced critical density of water,
$\rho_{wc}/\rho_{wt}=0.32207$. In this way, the resulting value of
the inverse range parameter is $z=1.7118$ which yields
$\bar{T}_{Yc}=1.35997$ and $\bar{\rho}_{Yc}=0.3019$. Since
$\bar{T}_{Yc}= k_B T_c/\epsilon_c$ (where we have added a subscript
c to $\epsilon$ for reasons that will become clear later) and
$\bar{\rho}_{Yt}=\rho_{wt} \sigma^3$, we get $\epsilon_c\simeq 6.57
\times 10^{-21}$ J and $\sigma\simeq 3.0401 \times 10^{-10}$ m.
Therefore we now have all the quantities required for the comparison
of the two coexistence curves. In figure \ref{fig1} we show the
liquid-vapor coexistence curves in the $T^*$-$\rho^*$ plane for both
real water and the corresponding attractive HCY fluid with the above
value of $z$.

Note that the shape of the two curves is similar. Nevertheless, one
still has to manipulate the data if a better quantitative agreement
is desired. Undoubtedly, the best way would be to follow a crossover
treatment\cite{crossover} incorporating the scaling laws valid for
the critical region, as has been done for instance in the case of
many compounds including water by Llovel \emph{et al}\cite{Lourdes}
in connection with a generalized van der Waals-type equation of
state. We, however, have taken a much simpler (pragmatic) approach
which suffices for our purposes and is described in what follows.

\subsection{Nonlinear rescaling for the temperature}

\begin{figure}[t]
\begin{center}
\includegraphics[height=7truecm]{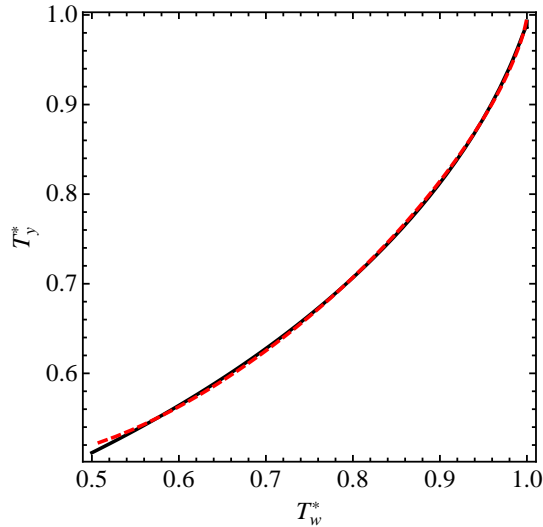}
\caption{Reduced temperature of the attractive HCY fluid as a
function of the reduced temperature of water in the saturation
branch of the vapor-liquid coexistence curve. The red dashed line
corresponds to data extracted from figure \ref{fig1} and the
continuous line corresponds to Eq.\ (\ref{fitTy}). } \label{fig2}
\end{center}
\end{figure}

From the saturation branch of the coexistence curves in figure 
\ref{fig1}, we extracted the two different reduced temperatures
corresponding to a fixed density and produced a plot $T^*_w$ vs
$T^*_Y$. This yields

\begin{equation}
T^{*}_Y=F(T^*_{w}),
\label{TYTW}
\end{equation}
where a good fit of the data produced in this way is a non linear
function of the form
\begin{eqnarray}
F(T^*_{w})=1-[a_1(1- T_{w}^{*})^{*1/4} + a_2 (1- T_{w}^{*})^{1/3} +
\nonumber \\ a_3 (1- T_{w}^{*})^{1/2} + a_4(1- T_{w}^{*})],
\label{fitTy}
\end{eqnarray}
with $a_1=1.39076$, $a_2=-3.34647$, $a_3=3.22505$ and $a_4=
-0.610568$. Figure \ref{fig2} displays the data corresponding to
this fit.

\begin{figure}[t]
\begin{center}
\includegraphics[height=7truecm]{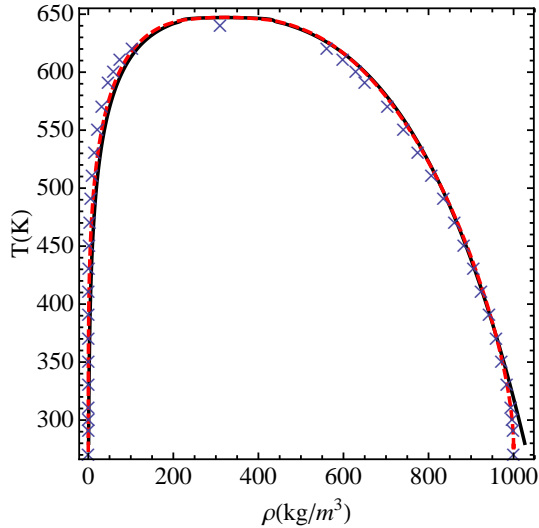}
\caption{Liquid-vapor coexistence curve as obtained with the recipe
(\ref{recipe}) (dashed line), compared to experimental
data\protect\cite{NIST} (continuous line) and the results of the
TIP4P/2005 model\protect\cite{Carlos2} (crosses).}
 \label{fig3}
\end{center}
\end{figure}

Now we address the question of whether the above information, namely
Eq.\ (\ref{fitTy}), allows us to `predict' the coexistence curve of
real water in real units using the results of the HCY fluid shown in
figure \ref{fig1}. The answer proceeds as follows. For any
temperature $T_w$ on the real water vapor-liquid equilibrium curve
we first determine the corresponding $T^*_w$ and then use Eq.\
(\ref{fitTy}) to obtain the corresponding $T^*_Y$. Next we multiply
it by $T_{Yc}$ and derive the corresponding vaporization and
condensation densities $\rho_{Yvap}$ and $\rho_{Ycond}$. Finally the
real densities are simply $\rho_{wt} \rho_{Yvap} /\rho_{Yt}$ and
$\rho_{wt}\rho_{Ycond}/\rho_{Yt}$. Schematically if we denote by
$f(T_Y,z)$ the mapping which allows one to determine ($\rho_{Yvap}$,
$\rho_{Ycond}$) in our approximation for the attractive HCY fluid,
then the recipe to obtain the real water liquid-vapor equilibrium
curve is:
\begin{equation}
(\rho_{wvap}, \rho_{wcond})=\frac{\rho_{wt}}{\rho_{Yt}}
f(F(T_w/T_{wc}) T_{Yc},1.7118).\label{recipe}
\end{equation}
Figure \ref{fig3} shows the predicted coexistence curve using this
recipe and the real water coexistence curve. In this figure we have
also included, for comparison, the results obtained from the
TIP4P/2005 model.\cite{Carlos2}

\begin{figure}[h]
\begin{center}
\includegraphics[height=7truecm]{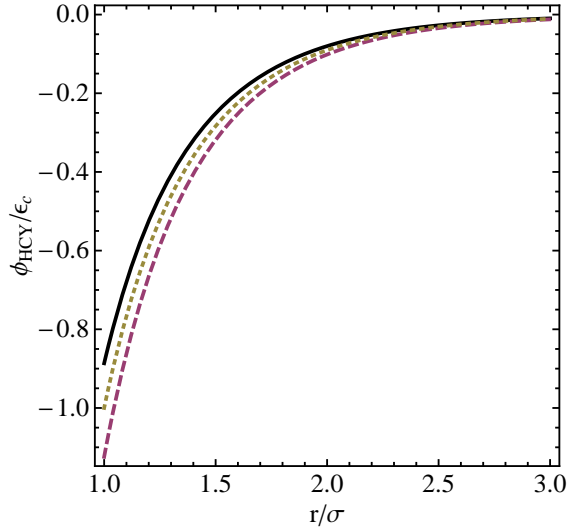}
\caption{Effective (temperature dependent) attractive HCY potential
as a function of distance for different temperatures: Continuous line: $T=273\ K$; dashed line: $473\ K$; dotted line:
$647\ K$.}
\label{fig3b}
\end{center}
\end{figure}

\begin{figure}[h]
\begin{center}
\includegraphics[height=7truecm]{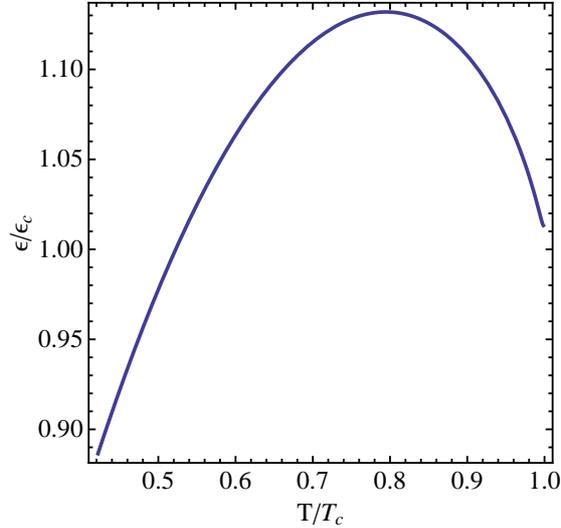}
\caption{Temperature dependence of the well depth $\epsilon (T)$ as
given by Eq.\ (\ref{epsilonTbis}).}
\label{fig3c}
\end{center}
\end{figure}

One can readily see the very good agreement between our prediction
and the real coexistence curve, except close to the triple point.  A
particulary noteworthy feature is that Eq.\ (\ref{fitTy}) was
derived by considering only the saturation branch, but it does a
very good job also for the rest of the coexistence curve. Moreover,
Eq.\ (\ref{fitTy}) admits the following appealing physical
interpretation. Assume that, instead of considering a constant depth
of the potential well at $r=\sigma$ (as we have done so far with
$\epsilon=\epsilon_c$), in order to make the HCY fluid
thermodynamically equivalent to real water for each temperature $T$
we are taking an effective HCY potential with the same $\sigma$ and
$z$ but with $\epsilon\equiv \epsilon (T)$. Then, Eq.\ (\ref{TYTW})
can be rewritten as
\begin{equation}
\frac{T/\epsilon(T)}{T_c/\epsilon_c}= F\left(\frac{T}{T_c}\right).
\label{epsilonT}
\end{equation}
or, equivalently,

\begin{equation}
\epsilon(T)=\epsilon_c \frac{T}{T_c F(T/T_c)}
\label{epsilonTbis}
\end{equation}
In figures \ref{fig3b} and \ref{fig3c} we show  the effective
(temperature dependent) potential as a function of distance and the
temperature dependence of its depth at $r=\sigma$, respectively.
Note that the overall shape of the potential is preserved, but the
effect of the extra (non monotonous) temperature dependence of
$\epsilon$ on the coexistence curve is clearly important. Although
such a task lies beyond the scope of the present paper, we are
persuaded that it would be interesting to investigate whether this
effective interaction may provide an adequate picture of the real
interaction between water molecules.

The next question that comes to mind is whether the same scaling of
the temperature and density will hold for the equation of state.
This is examined in the next subsection.

\subsection{The equation of state of liquid water}

The explicit form for our approximation to the attractive HCY fluid
equation of state is given in Eq.\ (\ref{PHCY}) but, with the above
discussion, it will be understood that we shall use in it the
temperature dependent $\epsilon(T)$ as given by Eq.\
(\ref{epsilonT}). However, one finds that the critical pressure
obtained in this way with the attractive HCY fluid is $P_{Yc}=38.08$
MPa whereas the real critical pressure is $P_{wc}=22.064$ MPa. This
overestimation seems to be due to a similar feature observed for the
results derived within the thermodynamic perturbation theory for the
attractive HCY fluid. In fact, in reference \cite{JPC2007} we
pointed out that the predicted critical pressure for the attractive
HCY fluid was always greater than simulation values. For instance,
if $z=1.8$ (which is the value of the inverse range parameter
closest to our $z=1.7118$ here where simulation results are
available), the critical pressure obtained from Eq. (\ref{PHCY})
is $\bar{P}=0.1576$ whereas from simulations one gets approximately $\bar{P}
=0.101$ \cite{Shukla}. This gives a ratio $0.101/0.1576=0.6409$. In
the present case, the ratio $P_{wc}/P_{Yc}$ yields
$\gamma=22.064/38.08=0.5794$ which could be reasonably attributed to
the discrepancy between the predicted critical pressure of the
attractive HCY fluid and the value that one would obtain from
simulation $z=1.7118$. Hence, we `correct' our equation of state
Eq.\ (\ref{PHCY}) with a factor $\gamma=22.064/38.08=0.5794$ so that
it now reads

\begin{equation}
P= \gamma \left[ \rho  k_B T Z_{{HS}}(\frac{\pi}{6}\rho \sigma^3)- 2
\pi {\rho}^2 \sigma^3 \epsilon(T) e^z \frac{\partial}{\partial \rho}
\left( \rho G(z) \right)\right].
\label{PHCY3}
\end{equation}

\begin{figure}[t]
\begin{center}
\includegraphics[height=7truecm]{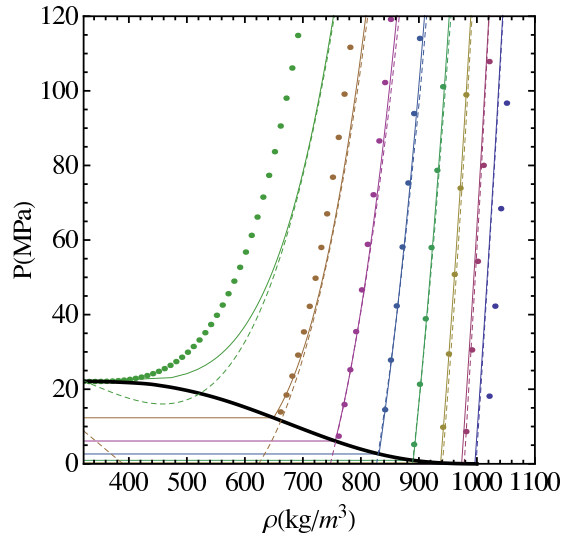}
\caption{Isotherms as obtained from experimental
data\protect\cite{NIST} (continuous lines), from Eq.
(\ref{PHCY3}) (dots) and from the Jeffery-Austin equation of
state\protect\cite{Jeffery} (dashed lines). From right to left the
curves correspond to $T=300$ K, $T=350$ K, $T=400$ K, $T=450$ K,
$T=500$ K, $T=550$, K $T=600$ K and $T=T_{wc}$. The real coexistence
line has been drawn to clearly indicate the liquid phase region.}
\label{figP2}
\end{center}
\end{figure}

Figure \ref{figP2} shows the result of using Eq.\ (\ref{PHCY3})
together with the experimental isotherms\cite{NIST} and those that
follow from the Jeffery-Austin equation of state \cite{Jeffery}.
Clearly the agreement between this corrected equation of state and
the experimental data for the isotherms is reasonably good in the
liquid region except close to the critical and triple point
isotherms. In the case of the critical isotherm, it leads to the
correct critical point but deviates from the real critical isotherm
at higher densities. On the other hand, the Jeffery-Austin results
are also remarkably accurate, but show similar limitations. In
particular they fail to produce the correct critical isotherm.

\section{Vapor-liquid equilibrium lines in the $P-T$ and $P-\rho$ planes.}
\label{sec3}
\begin{figure}[t]
\begin{center}
\includegraphics[height=7truecm]{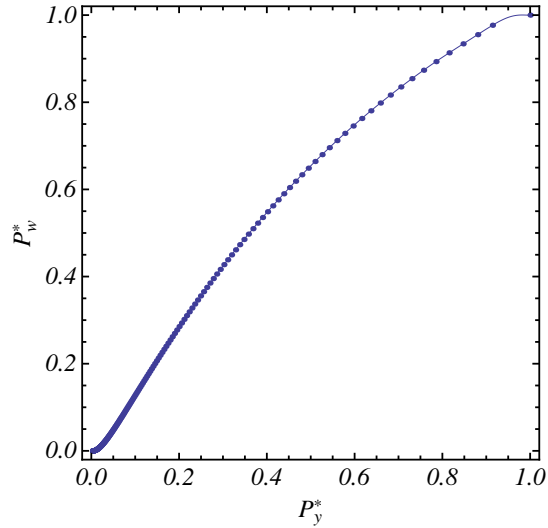} 
\caption{Relationship between the reduced pressures of the real water and of the HCY fluid as
derived from the coexistence curves in the $P-T$ plane. The continuous line is the fit given
in Eq. (\ref{PPRT})} \label{figPP}
\end{center}
\end{figure}

Once we have shown that the real vapor-liquid equilibrium
coexistence curve may be accurately  described in the $T$-$\rho$
plane through the use of our approximate (corrected) equation of
state, Eq.\ (\ref{PHCY3}), it is reasonable to wonder if using the
same approximate equation of state one may get also an accurate
description in the $P-T$ and $P-\rho$ planes. The answer to the
above question is only partially affirmative in the sense that the
approximation is very rough. In order to get an accurate description one is forced to make a subsequent
rescaling on the pressure which, in this instance, is unfortunately not simple due to
the form of the relationship between the reduced real pressure ($P_w^*=P_w/P_{wc}$) and the
reduced HCY pressure ($P^*_y=\bar{P}_y/\bar{P}_{yc}$) derived from the coexistence curves in the
$P-T$ plane and given by the curve in figure \ref{figPP}.  

In order to fit accurately this curve both in the low pressure end and close to the critical
pressure it is unavoidable to introduce a high order polynomial. We have taken the following form:
\begin{equation}
 P^{'}(P)=\sum_{i=1}^{14} c_i P^{i},
\label{PPRT}
\end{equation}
where the coefficients $c_i$ are given by $c_1=-0.149421$, $c_2=1.91855$, $c_3=-1.23117$, $c_4=0.519336$, $c_5=-0.150666$, $c_6=0.0306816$, $c_7=-0.004456$, $c_8=0.000466$, $c_9=-0.000035$, $c_{10}=1.880416\times10^{-6}$, $c_{11}=-7.003285\times10^{-8}$, $c_{12}=1.719548\times10^{-9}$, $c_{13}=-2.501064\times10^{-11}$ and $c_{14}=1.631257\times10^{-13}$.

\begin{figure}[t]
\begin{center}
\includegraphics[height=7truecm]{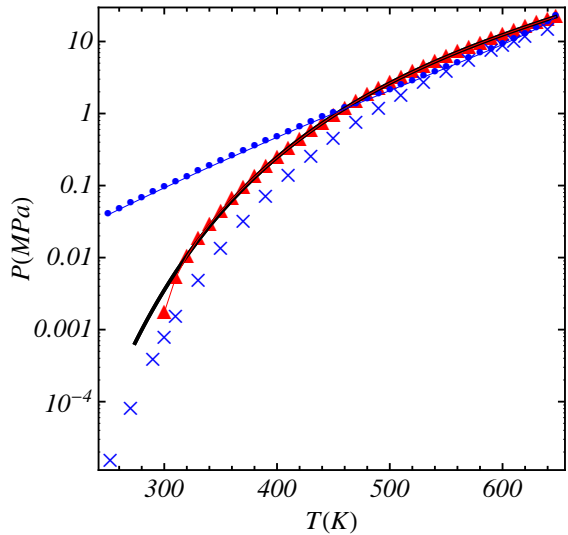}
\caption{Coexistence curves in the $P$ -$T$ plane in a log-linear
plot. Continuous line: experimental data\protect\cite{NIST}; dots:
attractive HCY fluid with temperature rescaled using Eq.
(\ref{fitTy}); triangles: attractive HCY fluid with temperature
rescaled using Eq. (\ref{fitTy}) and pressure rescaled using
Eq. (\ref{PPRT}); crosses: results from the TPI4P/2005
model\protect\cite{Carlos2}.}
\label{figPh1}
\end{center}
\end{figure}

In figure \ref{figPh1}, we show in a
log-linear plot the comparison between the different coexistence
curves in the $P$-$T$ plane. Note the disagreement between the
experimental curve and the one obtained with the attractive HCY
fluid using only Eq.\ (\ref{PHCY3}) and the substantial improvement
once the scaling implied by Eq. (\ref{PPRT}) is introduced. The inclusion of the results
obtained with the TPI4P/2005 model\cite{Carlos2} serves to indicate
that the qualitative tendency is reproduced by this model but the
pressure is underestimated.

\begin{figure}[t]
\begin{center}
\includegraphics[height=7truecm]{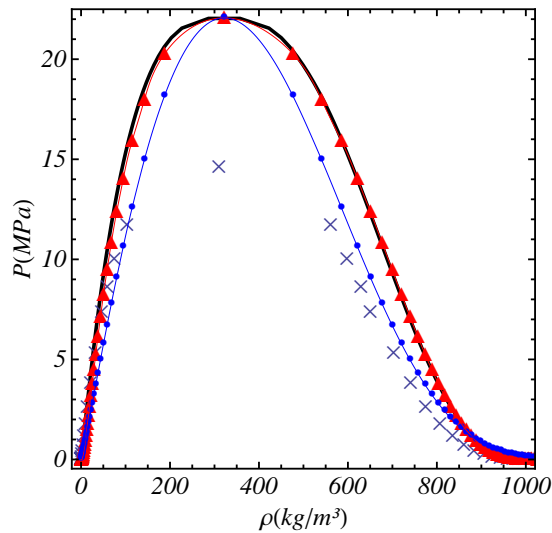} 
\caption{Coexistence curves in the $P$ - $\rho$ plane. Continuous line:
experimental data\protect\cite{NIST}; dots: attractive HCY fluid
with temperature rescaled using Eq. (\ref{fitTy}); triangles:
attractive HCY fluid with temperature rescaled using Eq.
(\ref{fitTy}) and pressure rescaled using Eq. (\ref{PPRT});
crosses: results from the TPI4P/2005 model\protect\cite{Carlos2}.}
\label{figPh2}
\end{center}
\end{figure}

Finally, we consider the $P-\rho$ plane. Figure \ref{figPh2}
contains the comparison between the different coexistence curves in
this plane. Note once more the disagreement between the experimental
curve and the one obtained with the attractive HCY fluid using Eq.\
(\ref{PHCY3}) and the improvement
when the scaling given by Eq. (\ref{PPRT}) is also used. Again, the TPI4P/2005 model\cite{Carlos0,Carlos2}
correctly reproduces the qualitative trends but in particular the
critical pressure is highly underestimated.

\section{Concluding remarks.}
\label{sec4}
As the results of the previous section indicate, we have been able
to use a simple model of intermolecular interaction, the attractive
HCY fluid, to obtain a very reasonable picture of the thermodynamic
behavior of water in the stable liquid phase. This is remarkable
given the fact that the model does not consider at all important
aspects such as bond geometry or charge distribution. In fact, our
formulation involving an effective (temperature-dependent)
attractive HCY interaction shares some aspects of the approach
followed in colloidal systems, where some degrees of freedom are
accounted for by an effective interaction potential. Moreover, the
present development is yet another effort in devising a sound
molecular theory of liquid water based on the principles of
statistical mechanics and the thermodynamic perturbation theory of
liquids. It is fair to say that with much less effort compared to
the IAWPS Formulation 1995\cite{Wagner} and probably also to the
Jeffery-Austin equation of state\cite{Jeffery} we are able to
achieve good accuracy. We should of course acknowledge the
semiphenomenological character of our approach that allowed us to
have this description, namely the fit of the inverse range parameter
of the model $z$ and the manipulations involved in the various
rescalings. Also worth mentioning in this regard is the limitation
to the liquid phase, which becomes manifest in that the agreement
between experimental data and our formulation gets worse both near
the critical point and near the triple point. 
Furthermore, the scaling implied in Eq. (\ref{fitTy})
 prevents us from dealing with the
supercritical region which is of course interesting in many industrial
processes. However, a similar procedure could be devised if one wanted to
deal with the equation of state of steam. In such a case, instead of rescaling
the temperature as we have done here, one could rescale the density. This might
also lead to a density-dependent pair potential which could be analyzed following
the recent approach introduced by Zhou \cite{Zhou}, although the use of density-dependent 
pair potentials is not free from controversy \cite{tejero}.
 Such developments lie beyond
the scope of the present paper.
Nevertheless, we seem
to have been able to capture the main physics behind the
thermodynamic properties of liquid water over a wide range of both
temperatures and pressures. Therefore, and given the fact that the
value of $z=1.7118$ that we used in our calculations is relatively
close to the one ($z\simeq 2$) in which the thermodynamic properties
of the attractive HCY are similar to those of the Lennard-Jones
fluid, we wonder whether the replacement of the Lennard-Jones
centers by attractive HCY centers in models used in simulation such
as the TPI4P/2005 model\cite{Carlos0} might improve the prediction
say of the critical pressure. This may only be ascertained by
performing the simulation and our hope is that, apart from the
potential use by engineers of the above results, this work may find
an echo in this latter respect.

\noindent \textbf{Acknowledgments}

We acknowledge the financial support of DGAPA-UNAM through project
IN -109408-2. The work of M.L.H. has also been supported by the
Ministerio de E- ducaci\'on y Ciencia (Spain) through Grant No.
FIS2007-60977 (partially financed by Feder funds) and by the Junta
de Extremadura through Grant No. GRU09038. We also want to thank A.
Santos, C. Vega and J. V. Sengers for helpful comments.

\bibliographystyle{elsarticle-num}

\begin{thebibliography}{00}
\bibitem{Carlos1} In the approximation for the equation of state of the attractive HCY fluid that we use here, the density
at the triple point coincides with the freezing density of the HS
fluid. The value we are using for this freezing density was taken
from E. G. Noya, C. Vega, and E. de Miguel, J. Chem. Phys. {\bf
128}, 154507 (2008).
\end{thebibliography}

\end{document}